\documentclass[twocolumn,showpacs,preprintnumbers,amsmath,amssymb,floatfix]{revtex4-1}
\usepackage{graphicx}
\usepackage{dcolumn}
\usepackage{isotope}
\usepackage{color}
\usepackage{bm}

\newcommand{\ve}[1]{\ensuremath{\mathbf{#1}}}
\newcommand{\n}[1]{\ensuremath{|\mathbf{#1}|}}
\newcommand{\Ep}{\ensuremath{E_N}}
\newcommand{\Epp}{\ensuremath{E_N'}}

\def\lsim{\lesssim}
\def\gsim{\gtrsim}

\begin{document}


\title{Analysis of $\gamma$-ray production in neutral-current neutrino-oxygen
interactions\\ at energies above 200 MeV}
\author{Artur M. Ankowski}
\altaffiliation[On leave from ]{Institute of Theoretical Physics, University of Wroc{\l}aw, Wroc{\l}aw, Poland} \email{Artur.Ankowski@roma1.infn.it}
\affiliation{INFN and Department of Physics,``Sapienza'' Universit\`a di Roma, I-00185 Roma, Italy}
\author{Omar Benhar}
\email{Omar.Benhar@roma1.infn.it}
\affiliation{INFN and Department of Physics,``Sapienza'' Universit\`a di Roma, I-00185 Roma, Italy}
\author{Takaaki Mori}
\affiliation{Department of Physics, Okayama University, Okayama 700-8530, Japan}
\author{Ryuta Yamaguchi}
\affiliation{Department of Physics, Okayama University, Okayama 700-8530, Japan}
\author{Makoto Sakuda}
\email{sakuda@fphy.hep.okayama-u.ac.jp}
\affiliation{Department of Physics, Okayama University, Okayama 700-8530, Japan}

\date{\today}%

\begin{abstract}
It has long been recognized that observation of the $\gamma$ rays originating from nuclear deexcitation can be
exploited to identify neutral-current neutrino-nucleus interactions in water-Cherenkov detectors.
We report the results of a calculation of the neutrino- and antineutrino-induced $\gamma$-ray production cross section
for the oxygen target. Our analysis is focused on the kinematical region of neutrino energy larger than $\sim$200 MeV, in which a~single-nucleon
knockout is known to be the dominant reaction mechanism. The numerical results have been obtained using for the first time
a realistic model of the target spectral function, extensively tested against electron-nucleus scattering data.
We find that at a~neutrino energy of 600~MeV the fraction of neutral-current interactions leading to emission of $\gamma$ rays of energy larger than 6~MeV is $\sim$41\%, and that
the contribution of the $p_{3/2}$ state is overwhelming.

\end{abstract}

\pacs{25.30.Pt, 21.10.Pc, 23.20.Lv}
%
%
%
%


\maketitle

The observation of $\gamma$ rays originating from nuclear deexcitation can be exploited to identify neutral-current (NC)
neutrino-nucleus interactions in a broad energy range. The authors of Ref.~\cite{KLV} first suggested to use this signal to detect supernova neutrinos,
the average energy of which is $\sim$25 MeV. Interactions of atmospheric neutrinos, with energies extending to the GeV region,
can also lead to transitions to excited nuclear states decaying through $\gamma$-ray emission,  possibly associated with a~hadronic cascade~\cite{nussinov}.

Neutrons, while providing  $\sim$50\% of NC events, do not emit Cherenkov light. As a consequence, the availability
of an alternative signal allowing one to identify NC interactions is very important. Events with $\gamma$ rays of energy above the observational threshold of  5~MeV can be detected in a water-Cherenkov detector,
like Super-Kamiokande, and contribute up to $\sim$5\% of the total event number~\cite{Beacom,T2K}, independent of neutrino oscillations. Note that in water $\sim$90\% (16 out of 18) of the NC interactions take place in oxygen.

Following the pioneering studies of nuclear excitations by neutral weak currents of Refs.~\cite{Donnelly,Donnelly&Peccei},
theoretical calculations of the cross section of $\gamma$-ray production from NC neutrino-oxygen interactions
have been carried out in the neutrino energy range $E_\nu\sim 10$--500 MeV~\cite{KLV,Kolbe1,Kolbe2}.
These studies took into account $\gamma$ rays originating from the inelastic processes
 $\nu+\isotope[16][8]{O}\rightarrow \nu^\prime + \isotope[16][8]{O}^*$,
 in which the oxygen nucleus is mainly excited
 to resonances lying above particle emission threshold. These states
 then decay to either $p+\isotope[15][7]{N}^*$ or $n+\isotope[15][8]{O}^*$, and
 the residual nuclei, left in excited particle-bound states, decay in turn emitting $\gamma$ rays
 in the 5--10~MeV region.

At low energy, elastic scattering and inelastic excitation of discrete nuclear states provide the main contribution to
the neutrino-nucleus cross section. However,  at $E_\nu \gsim 200$~MeV the cross section associated with these processes
tends to saturate, and quasielastic (QE) nucleon knockout becomes the dominant reaction mechanism. If the residual nucleus is
left in an excited state, these processes can also lead to $\gamma$-ray emission.
The K2K Collaboration reported the observation of $\gamma$ rays from nuclear deexcitation
following NC neutrino-oxygen interactions  at $E_{\nu}\sim1.3$~GeV
in the 1-kton water-Cherenkov detector~\cite{Kameda}. The
number of events and the visible energy are qualitatively consistent with those expected from
6-MeV $\gamma$-ray production in NC QE neutrino-oxygen interactions.

\begin{figure}
\includegraphics[width=0.60\columnwidth]{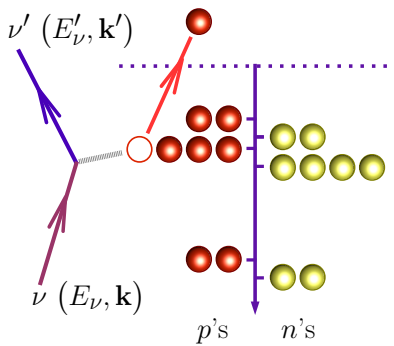}
\caption{\label{fig1} (color online). Schematic representation of neutral-current neutrino scattering off oxygen.}
\end{figure}

In the QE regime, neutrino-nucleus scattering reduces to the incoherent sum of
elementary scattering processes involving individual nucleons, the energy and momentum of which are distributed according to the
target spectral function~\cite{RMP}. A schematic representation of NC QE neutrino-nucleus scattering is given in Fig.~\ref{fig1}, where the dashed line
represents the threshold for nucleon emission in the continuum.

In this Letter, we discuss the emission of $\gamma$ rays
arising from the decay of the residual nuclei of the reactions
$\nu+\isotope[16][8]{O}\rightarrow \nu+p+\isotope[15][7]{N}^*$ or $\nu+\isotope[16][8]{O}\rightarrow \nu+n+\isotope[15][8]{O}^*$,
the cross sections of which have been computed using a realistic model of the oxygen spectral function.

Note that, due to the strong energy-momentum correlation exhibited
by the nuclear spectral function, large excitation energies of the
residual system are associated with large momenta of the
knocked out nucleon. As nucleons occupying shell-model states have
a vanishingly small probability of carrying momentum larger than
$\sim$250 MeV \cite{PDS}, knockout of these nucleons predominantly leaves the
residual system in a bound state.

\begin{figure}
\includegraphics[width=0.80\columnwidth]{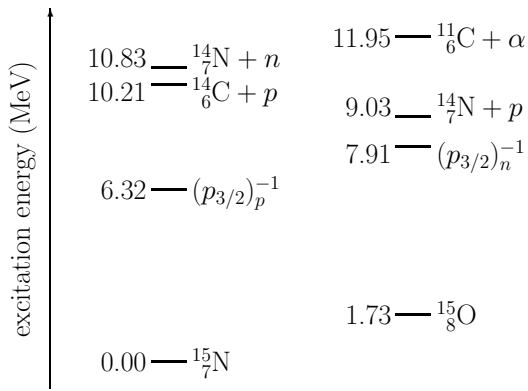}
\caption{\label{fig2}
Low-lying excited levels of the residual nuclei produced in $\isotope[16][8]{O}(\nu,\nu'N)$ scattering. Energies are measured with respect to the \isotope[15][7]{N} ground state.
}
\end{figure}

In our approach, the cross section of $\gamma$-ray production following a NC QE interaction, $\sigma_\gamma$,  is written in the form
\begin{eqnarray}\label{eq:gammaCS}
& & \sigma_\gamma\equiv\sigma(\nu + \isotope[16][8]{O} \to \nu + \gamma +Y +N)\\
\nonumber
&  &  = \sum_\alpha\sigma(\nu + \isotope[16][8]{O} \to \nu+X_\alpha + N )\:\textrm{Br}(X_\alpha \to \gamma +Y),
\end{eqnarray}
where $N$ is the knocked out nucleon, $X_\alpha$ denotes the residual nucleus in the state $\alpha$,
and $Y$ is the system resulting from the electromagnetic decay of $X_\alpha$, e.g. \isotope[15][8]{O}, \isotope[15][7]{N},
$\isotope[14][7]{N} +n$, or $\isotope[14][6]{C} +p$~\cite{Kamyshkov, Isotopes1, Isotopes2}. The energy spectrum of the states of the
residual nuclei is schematically illustrated in Fig.~\ref{fig2}.

According to the shell model, nuclear dynamics can be described by a mean field.
In the simplest implementation of this model, protons in the $\isotope[16][8]{O}$ nucleus occupy three states, $1p_{1/2}$, $1p_{3/2}$, and $1s_{1/2}$, with
removal energy 12.1, 18.4, and $\sim$42~MeV, respectively~\cite{Saclay,Nikhef,JLab}. The neutron levels exhibit the same pattern, see Fig.~\ref{fig1}, but are more deeply bound by 3.54~MeV~\cite{Isotopes2}. Since below nucleon-emission threshold the deexcitation process is governed only by energy differences,
the proton and neutron holes yield photons of very similar energy, the differences being as small as $\sim$0.1~MeV (see Fig.~\ref{fig2}).

The calculation of the NC QE cross section, $\sigma(\nu + \isotope[16][8]{O} \to \nu+X_\alpha + N)$, has been
performed  within the approach discussed in Refs.~\cite{Benhar1,Meloni} for the case of charged-current (CC) interactions, whereas the branching ratios
Br$(X_\alpha \to \gamma +Y)$  have been taken from Refs.~\cite{Ejiri,Kamyshkov}.

Following Refs.~\cite{Benhar1,Meloni}, we write the NC QE cross section in the form
\begin{eqnarray}
\frac{d\sigma_{\nu A}}{d\Omega dE'_\nu }= \sum_{N=p,\,n}\int d^3p\,
dEP_N({\bf p},E) \frac{M}{\Ep}\frac{d\sigma _{\nu N}}{d\Omega dE'_\nu },
\label{NC:xsec}
\end{eqnarray}
where $\Ep = \sqrt{M^2 + \ve p^2}$, $M$ being the nucleon mass,
$d\sigma _{\nu N}/d\Omega dE^\prime_\nu $ denotes the elementary neutrino-nucleon cross section and
the spectral function $P_N({\bf p},E)$ yields the probability of removing a nucleon of momentum ${\bf p}$ from the
target leaving the residual nucleus with energy $E + E_0 -M$, $E_0$ being the target ground-state energy.

In the nuclear shell model, nucleons occupy single-particle states $\phi_{\alpha}$ with binding
energy $-E_\alpha$ ($E_\alpha > 0$).
As a consequence, knockout of a target nucleon leaves the residual system in a bound state, and the spectral function can be
conveniently written in the form
\begin{equation}
P_N({\bf p},E) = \sum_{\alpha\: \in \{F\} } n_\alpha |\phi_\alpha({\bf p})|^2 f_\alpha (E-E_\alpha),
\label{S:MF}
\end{equation}
where $\phi_{\alpha}({\bf p})$ is the momentum-space wave function associated with the $\alpha$th
shell model state and the sum is extended to all occupied states belonging to the Fermi sea $\{F\}$.
The {\it occupation probability} $n_\alpha \leq 1$ and the (unit-normalized) function $f_\alpha(E-E_\alpha)$, describing
the energy width of the $\alpha$th state, account for the effects of nucleon-nucleon (NN) correlations,
not included in the mean-field picture.
In the absence of correlations, $n_\alpha \rightarrow 1$ and $f_\alpha(E-E_\alpha) \rightarrow \delta(E-E_\alpha)$.

Precise measurements of the coincidence $(e,e^\prime p)$ cross section, yielding direct access to the target spectral function, have
provided unambiguous evidence of deviations from the mean-field scenario, leading to significant depletion of the single- particle states~\cite{Saclay,Nikhef,JLab}. The data at large missing momentum and large missing energy [i.e. large $\n p$ and large $E$ in Eq.~\eqref{NC:xsec}],
 collected at Jefferson Lab  by the JLAB E97-006 Collaboration, indicate that NN correlations
push $\sim$20\% of the total strength to continuum states outside the Fermi sea~\cite{Daniela}.

A realistic model of the proton spectral function of oxygen has been obtained within the
local density approximation (LDA), combining the experimental data of
Ref.~\cite{Saclay} with the results of theoretical calculations of the correlation contribution
in uniform nuclear matter at different densities \cite{Benhar2,Benhar1}.
The results reported in Ref.~\cite{Benhar1} show that the LDA spectral function provides an
accurate description of the inclusive electron-oxygen cross sections at beam energies around
1~GeV. In addition, it predicts a nucleon momentum distribution in agreement with that obtained from
the data of Ref.~\cite{Daniela}.

As pointed out by the authors of Ref.~\cite{Benhar_90}, nucleon-knockout experiments measure {\em spectroscopic strengths}, not occupation probabilities.
Spectroscopic strengths are given by the area below the sharp peaks observed in the missing-energy spectra, corresponding to knockout of a nucleon occupying one of the
shell-model states, corrected to take into account final-state interactions. On the other hand, occupation probabilities include contributions corresponding to
larger removal energy, arising from mixing of the one-hole state with more complex final states \cite{Benhar_90}.

\begin{figure}
\includegraphics[width=0.80\columnwidth]{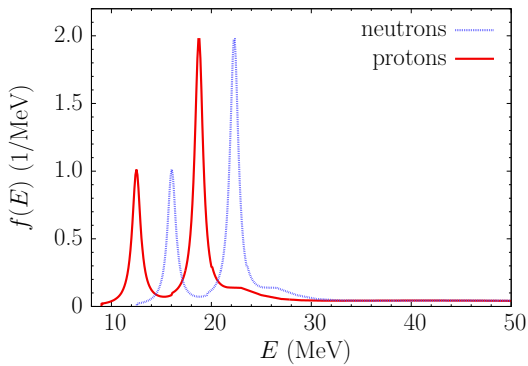}
\caption{\label{fig3}(color online). Distribution of removal energy of protons and neutrons in \isotope[16][8]{O}.}
\end{figure}

Figure~\ref{fig3} shows the energy distribution obtained from
momentum integration of the spectral function of Refs.~\cite{Benhar2,Benhar1}. It clearly appears that, unlike the  $p_{1/2}$ and $p_{3/2}$ states,
the $s_{1/2}$ state is spread out over a broad energy range,
and can hardly be treated as a single-particle state.

The $p_{1/2}$,  $p_{3/2}$ and $s_{1/2}$ spectroscopic strengths have been computed by integrating
the oxygen spectral function of Refs.~\cite{Benhar2,Benhar1} over the energy ranges
$11.0 \leq E \leq 14.0$ MeV, $17.25 \leq E \leq 22.75$ MeV, and $22.75 \leq E \leq 62.25$ MeV, respectively.
Dividing these numbers by the degeneracy of the shell-model states, one obtains the quantities $S_\alpha$  listed in Table~\ref{tab1}.
The same spectroscopic strengths have been used for protons and neutrons.

Our results turn out to be very close to those extracted from  the high resolution measurement carried out at NIKHEF-K~\cite{Nikhef}.
For example, the $p_{1/2}$ ($p_{3/2}$) strength collected in the same energy range is reported to be 0.630$\pm$0.034 (0.676$\pm$0.037).

The uncertainty in the determination of $S_\alpha$ is mainly due to  the choice of the shell-model wave functions and to the
treatment of final-state interactions of the knocked-out proton. The authors of Ref.~\cite{Nikhef} quote an overall systematic uncertainty
of 5.4\%.

\begin{table}
\caption{\label{tab1} Spectroscopic strengths of the \isotope[16][8]{O} hole states and their branching ratios for deexcitation by the $E_\gamma>6$ MeV photon emission.}
\begin{ruledtabular}
    \begin{tabular}{@{}l|lll@{}}
    $\alpha$  &  $p_{1/2}$  &  $p_{3/2}$ & $s_{1/2}$\\[3pt]
    \hline
    $S_\alpha$  & 0.632 & \phantom{00}0.703 & \phantom{0}0.422\\
    Br$(X_\alpha\to \gamma+Y)$  & 0\% & 100\% & $16\pm1$\%\\
    \end{tabular}
\end{ruledtabular}
\end{table}

The elementary neutrino-nucleon cross section of Eq.~\eqref{NC:xsec} can be written in the form
\begin{eqnarray}\label{eq:elementaryCS}
\frac{d^2 \sigma_{\nu N}}{d\Omega dE'_{\nu}}=\frac{G^2_F}{8 \pi^2}\frac{E'_\nu}{E_\nu}
\frac{L_{\mu \nu} W^{\mu \nu}}{M\Epp}\:\delta(\tilde\omega+\Ep-\Epp),
\end{eqnarray}
where $\Epp=\sqrt{M^2+\ve{p'}^2}$. The leptonic and hadronic tensor, $L_{\mu \nu}$ and $ W^{\mu \nu}$, are
given by
\begin{equation}\label{eq:leptonicT}
L_{\mu \nu} = 2\big(k'_\mu k_\nu + k'_\nu k_\mu -g_{\mu \nu}k \cdot k' -
i\varepsilon _{\mu \nu \alpha \beta}k^{\alpha} k'^{\beta}\big)
\end{equation}
and
\begin{equation}\begin{split}
W^{\mu \nu} &= -g^{\mu \nu}M^2 W_1 + \tilde p^{\mu}\tilde p^{\nu}W_2+i \varepsilon ^{\mu \nu \alpha \beta}
\tilde p_\alpha \tilde{q}_\beta W_3\\
&\quad + \tilde{q}^{\mu}\tilde{q}^{\nu}W_4 + (\tilde p^{\mu}\tilde{q}^{\nu} + \tilde p^{\nu}\tilde{q}
^{\mu})W_5,
\end{split}\end{equation}
with $\tilde p=(\Ep, \ve p)$ and $\tilde q=(\tilde\omega, \ve k - \ve{k'})$. As in the case of CC QE
scattering~\cite{Meloni}, the structure functions $W_i$ can be written in terms of the nucleon form factors according to
\begin{equation}\begin{split}
W_1&=\tau\big(\mathcal{F}_1^N+\mathcal{F}_2^N\big)^2 + (1 +\tau)\mathcal{F}_A^2, \\
W_2&=\big(\mathcal{F}_1^N\big)^2 + \tau\big(\mathcal{F}_2^N\big)^2+\mathcal{F}_A^2, \\
W_3&=\big(\mathcal{F}_1^N+\mathcal{F}_2^N\big)\mathcal{F}_A, \\
W_4&=\frac14\Big[\big(\mathcal{F}_1^N\big)^2 + \tau\big(\mathcal{F}_2^N\big)^2-\big(\mathcal{F}_1^N+\mathcal{F}_2^N\big)^2 \\
&\quad-4\mathcal{F}_p\big(\mathcal{F}_A-\tau\mathcal{F}_p\big)\Big],\\
W_5&=\frac{1}{2}W_2,
\end{split}\end{equation}
with $\tau=-\tilde q^2/(4M^2)$.
Note that, in the above equations, the electromagnetic and charged-current nucleon form factors
$\{ F^N_i\ (i=1,2), F_A, F_p \}$ are replaced by the ones appropriate to describe NC interactions~\cite{Weinberg,Donnelly&Peccei,Alberico}
\begin{equation}\begin{split}
\mathcal{F}_i^N&=\pm \frac{1}{2}(F_i^p-F_i^n)-2\sin^2\theta_W F_i^N,\\
\mathcal{F}_A&=\frac{1}{2}\big(F_A^s\pm F_A\big)=\frac{1}{2}\frac{\Delta s\pm g_A}{(1-\tilde q^2/M^2_A)^2},\\
\mathcal{F}_p&=\frac{2M^2\mathcal{F}_A}{m^2_\pi -\tilde q^2},
\end{split}\end{equation}
where the upper (lower) sign corresponds to proton (neutron) form factors, $\theta _W$ is the weak mixing angle, $m_\pi$ is the pion mass, $g_A=-1.2673$, and the strange quark contribution is set to
$\Delta s=-0.08$~\cite{strange}. The form factors
$F_1^N$ and $F_2^N$ can be expressed in terms of the measured Sachs form factors
$G^N_E$ and $G^N_M$ as
\begin{equation}\begin{split}
F_1^N = \frac{G^N_E+\tau G^N_M}{1+\tau},\qquad F_2^N = \frac{G^N_M-G^N_E}{1+\tau}.
\end{split}\end{equation}
In this Letter, we use the state-of-the-art parametrization of $G^N_E$ and $G^N_M$ of Ref.~\cite{FormFactors}.

The branching ratios $\textrm{Br}(X_\alpha \to \gamma +Y)$, necessary to calculate the cross section $\sigma_\gamma$ according to
Eq.~\eqref{eq:gammaCS}, are collected in Table~\ref{tab1}~\cite{Ejiri, Kamyshkov}. In the case of the $p_{1/2}$-proton (neutron) knockout, the residual nucleus is \isotope[15][7]{N} (\isotope[15][8]{O}) produced in its ground state. Hence, no $\gamma$ rays are produced. As the $p_{3/2}$-proton (neutron) hole lies below the nucleon-emission threshold,  10.21~MeV (7.30~MeV), it always deexcites through photon emission with half-life $0.146\pm0.008$ fs (less than 1.74 fs)~\cite{Isotopes2}. When a proton (neutron) is knocked out from the deepest $s_{1/2}$ shell, the excitation energy is high enough for many deexcitation channels to open, of which only two, $\isotope[14][6]{C}+p$ and $\isotope[14][7]{N}+n$ ($\isotope[14][6]{C}+p$ and $\isotope[11][6]{C}+\alpha$), yield photons of energy higher than 6~MeV~\cite{Kamyshkov} (see Fig.~\ref{fig2}). The theoretical estimate of the branching ratio for these processes~\cite{Ejiri, Kamyshkov}, being in total 16\%, turns out to be in good agreement with the value $15.6\pm 1.3^{+0.6}_{-1.0}\%$
extracted from the $\isotope[16][8]{O}(p,2p)\isotope[15][7]{N}$ experiment E148 carried out at the Research Center for Nuclear Physics (RCNP) of the Osaka
University~\cite{E148}.

Although the $s_{1/2}$ contribution to $\sigma_\gamma$ is an increasing function of neutrino energy, it saturates at $\sim$0.4~GeV and remains less than 4.5\% for $E_\nu\leq5$~GeV. Compared to the $p_{3/2}$ contribution, it is suppressed by the low branching ratio for deexcitation through $\gamma$ emission, the lower strength
and degeneracy,  as well as the larger removal energy, that makes the NC QE cross section smaller.

Note that the formalism presented in this Letter also applies to antineutrino-induced $\gamma$-ray production, the only difference being the sign of the last term in Eq.~\eqref{eq:leptonicT}.

\begin{figure}
\includegraphics[width=0.80\columnwidth]{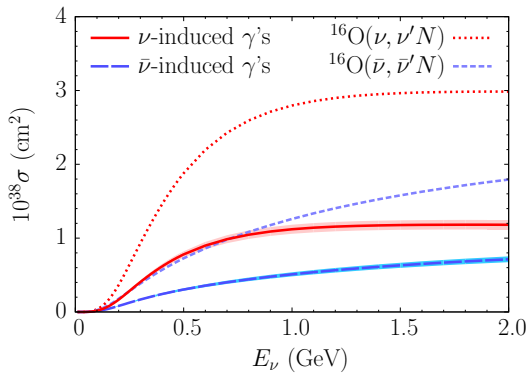}
\caption{\label{fig4} (color online). Cross section for $\gamma$-ray production following NC QE interaction of neutrino (solid line) and antineutrino (long-dashed line) compared to the NC QE cross section of neutrino (dotted line) and antineutrino (short-dashed line). Only the photons of energy larger that 6~MeV are considered.
}
\end{figure}

In Fig.~\ref{fig4} our results for the neutrino- and antineutrino-induced $\gamma$-ray production cross section are compared to the
neutrino and antineutrino NC QE cross sections. The error bands show the uncertainties arising form the determination of the spectroscopic strengths (5.4\%),
the treatment of Pauli blocking (1\%), and the branching ratio of the $s_{1/2}$ state (1\%).
The $\sigma_\gamma$'s dependence on neutrino energy is very similar, although not identical, to that of the NC QE cross section. The discrepancy arises from
 difference between the average removal energy associated with the whole spectral function and the energy of the $p_{3/2}$ shell, yielding the overwhelming
 contribution to $\sigma_\gamma$. The neutrino-induced $\gamma$-production cross section reaches its maximum at $E_\nu\sim1.9$ GeV and is slowly decreasing at larger energies. On the other hand, the corresponding antineutrino cross section is an increasing function of $E_\nu$.

As the axial mass enters $\sigma_\gamma$ only through the elementary cross section~\eqref{eq:elementaryCS}, the ratio $\sigma_\gamma/\sigma_\textrm{NC}$, $\sigma_\textrm{NC}$ being the NC QE cross section, is largely independent of $M_A$.  For example, applying $M_A=1.39$ GeV (1.03 GeV) instead of the value 1.2~GeV used in this Letter~\cite{Gran}, changes the ratio $\sigma_\gamma/\sigma_\textrm{NC}$ by less than 0.4\% (0.3\%) when $E_\nu\leq5$~GeV.

The mechanism of $\gamma$-ray production through nuclear deexcitation is the same for NC and CC processes. Therefore, the fraction of
neutrino interactions followed by $\gamma$-ray emission in the two cases is determined by the same factors.
In the case of CC QE scattering, the maximum value of the energy transfer is reduced by the nonvanishing mass of the charged lepton, and therefore the contribution
of the $p_{3/2}$ knockout to the total cross section is somewhat more significant. However, this effect is already small at  $E_\nu=475$~MeV, as the fraction of CC interactions emitting photons is higher than that of NC interactions by only 1\%, and becomes even smaller with increasing neutrino energy.

In conclusion, we have computed the neutrino and antineutrino neutral-current cross sections, focusing on the kinematical region in which
single-nucleon knockout dominates. In this region the average of neutrino
and antineutrino cross sections obtained from our approach is much larger
than the corresponding result of Ref. \cite{Kolbe2}. For example,
at $E_\nu = 0.5$ (1.0)~GeV our average cross section exceeds the
one reported in Ref.~\cite{Kolbe2} by a factor~$\sim$10 (15).

The NC cross sections have been used to compute the $\gamma$-ray production
cross sections. Considering photons of energy larger than 6~MeV, we find that
the $p_{3/2}$ state provides the overwhelming contribution, and that the
ratio $\sigma_\gamma/\sigma_{{\rm NC}}$, exhibiting a significant
energy-dependence at $E_\nu \lsim 1$ GeV, is $\sim$41\% at $E_\nu=600~$MeV.

Our results,
obtained using a realistic model of the target spectral function, provide an accurate estimate of a signal that
can be exploited to identify neutral-current events in water-Cherenkov detectors.

This work was supported by INFN (Grant MB31), MIUR PRIN (Grant ``Many-body theory of nuclear systems and implications on the
physics of neutron stars''), and in part by the Grant-In-Aid from the Japan Society for Promotion of Science (Nos. 21224004 and 23340073).
A.M.A. was supported by the Polish Ministry of Science and Higher Education under Grant No. 550/MOB/2009/0.

\end{document}